\PassOptionsToPackage{unicode}{hyperref}
\PassOptionsToPackage{hyphens}{url}
\PassOptionsToPackage{dvipsnames,svgnames,x11names}{xcolor}
\documentclass[
  12pt]{article}

\usepackage{amsmath,amssymb}
\usepackage{lmodern}
\usepackage{iftex}
\ifPDFTeX
  \usepackage[T1]{fontenc}
  \usepackage[utf8]{inputenc}
  \usepackage{textcomp} 
\else 
  \usepackage{unicode-math}
  \defaultfontfeatures{Scale=MatchLowercase}
  \defaultfontfeatures[\rmfamily]{Ligatures=TeX,Scale=1}
\fi
\IfFileExists{upquote.sty}{\usepackage{upquote}}{}
\IfFileExists{microtype.sty}{
  \usepackage[]{microtype}
  \UseMicrotypeSet[protrusion]{basicmath} 
}{}
\makeatletter
\@ifundefined{KOMAClassName}{
  \IfFileExists{parskip.sty}{%
    \usepackage{parskip}
  }{
    \setlength{\parindent}{0pt}
    \setlength{\parskip}{6pt plus 2pt minus 1pt}}
}{
  \KOMAoptions{parskip=half}}
\makeatother
\usepackage{xcolor}
\ifx\paragraph\undefined\else
  \let\oldparagraph\paragraph
  \renewcommand{\paragraph}[1]{\oldparagraph{#1}\mbox{}}
\fi
\ifx\subparagraph\undefined\else
  \let\oldsubparagraph\subparagraph
  \renewcommand{\subparagraph}[1]{\oldsubparagraph{#1}\mbox{}}
\fi

\usepackage{longtable,booktabs,array}
\usepackage{calc} 
\usepackage{etoolbox}
\makeatletter
\patchcmd\longtable{\par}{\if@noskipsec\mbox{}\fi\par}{}{}
\makeatother
\IfFileExists{footnotehyper.sty}{\usepackage{footnotehyper}}{\usepackage{footnote}}
\makesavenoteenv{longtable}
\usepackage{graphicx}
\makeatletter
\def\maxwidth{\ifdim\Gin@nat@width>\linewidth\linewidth\else\Gin@nat@width\fi}
\def\maxheight{\ifdim\Gin@nat@height>\textheight\textheight\else\Gin@nat@height\fi}
\makeatother
\setkeys{Gin}{width=\maxwidth,height=\maxheight,keepaspectratio}
\makeatletter
\def\fps@figure{htbp}
\makeatother

\makeatletter
\@ifpackageloaded{caption}{}{\usepackage{caption}}
\AtBeginDocument{%
\ifdefined\contentsname
  \renewcommand*\contentsname{Table of contents}
\else
  \newcommand\contentsname{Table of contents}
\fi
\ifdefined\listfigurename
  \renewcommand*\listfigurename{List of Figures}
\else
  \newcommand\listfigurename{List of Figures}
\fi
\ifdefined\listtablename
  \renewcommand*\listtablename{List of Tables}
\else
  \newcommand\listtablename{List of Tables}
\fi
\ifdefined\figurename
  \renewcommand*\figurename{Figure}
\else
  \newcommand\figurename{Figure}
\fi
\ifdefined\tablename
  \renewcommand*\tablename{Table}
\else
  \newcommand\tablename{Table}
\fi
}
\@ifpackageloaded{float}{}{\usepackage{float}}
\floatstyle{ruled}
\@ifundefined{c@chapter}{\newfloat{codelisting}{h}{lop}}{\newfloat{codelisting}{h}{lop}[chapter]}
\floatname{codelisting}{Listing}

\makeatother
\makeatletter
\@ifpackageloaded{caption}{}{\usepackage{caption}}
\@ifpackageloaded{subcaption}{}{\usepackage{subcaption}}
\makeatother
\makeatletter
\@ifpackageloaded{tcolorbox}{}{\usepackage[many]{tcolorbox}}
\makeatother
\makeatletter
\@ifundefined{shadecolor}{\definecolor{shadecolor}{rgb}{.97, .97, .97}}
\makeatother
\ifLuaTeX
  \usepackage{selnolig}  
\fi
\usepackage{bm} 
\usepackage[]{natbib}
\usepackage[all]{xy} 
\IfFileExists{bookmark.sty}{\usepackage{bookmark}}{\usepackage{hyperref}}
\IfFileExists{xurl.sty}{\usepackage{xurl}}{} 
\hypersetup{
  pdftitle={Title},
  pdfkeywords={Neuroimaging; Early Life Adversity; Latent Variables; Predictive Modeling; Multi-omics Integration; Hierarchical Bayesian Models},
  colorlinks=true,
  linkcolor={blue},
  filecolor={Maroon},
  citecolor={Blue},
  urlcolor={Blue},
  pdfcreator={LaTeX via pandoc}}

\begin{document}

\def\spacingset#1{\renewcommand{\baselinestretch}{#1}\small\normalsize} \spacingset{1}


\title{\bf A Bayesian Integrative Mixed Modeling Framework for Analysis of the Adolescent Brain and Cognitive Development Study}
\author{
Aidan Neher\thanks{Correspondence concerning this article should be addressed to Aidan Neher at neher015@umn.edu}, Apostolos Stamenos, Mark Fiecas, Sandra E. Safo, Thierry Chekouo\thanks{or Thierry Chekouo at tchekouo@umn.edu.}
\\Biostatistics and Health Data Science, University of Minnesota\\
}
\maketitle

\bigskip
\bigskip
\begin{abstract}
Integrating high-dimensional, heterogeneous data from multi-site cohort studies with complex hierarchical structures poses significant feature selection and prediction challenges. We extend the Bayesian Integrative Analysis and Prediction (BIP) framework to enable simultaneous feature selection and outcome modeling in data of nested hierarchical structure. We apply the proposed Bayesian Integrative Mixed Modeling (BIPmixed) framework to the Adolescent Brain Cognitive Development (ABCD) Study, leveraging multi-view data, including structural and functional MRI and early life adversity (ELA) metrics, to identify relevant features and predict the behavioral outcome. 
BIPmixed incorporates 2-level nested random effects, to enhance interpretability and make predictions in hierarchical data settings. Simulation studies illustrate BIPmixed's robustness in distinct random effect settings, highlighting its use for complex study designs. Our findings suggest that BIPmixed effectively integrates multi-view data while accounting for nested sampling, making it a valuable tool for analyzing large-scale studies with hierarchical data.
\end{abstract}

\noindent%
\vfill

\newpage
\spacingset{1.8} 
\ifdefined\Shaded\renewenvironment{Shaded}{\begin{tcolorbox}[sharp corners, enhanced, breakable, borderline west={3pt}{0pt}{shadecolor}, interior hidden, boxrule=0pt, frame hidden]}{\end{tcolorbox}}\fi

\hypertarget{sec-intro}{%
\section{Introduction}\label{sec-intro}}

\label{s:intro}

ur objective in this paper is to use data from the Adolescent Brain and Cognitive Development (ABCD) Study, a multi-site cohort study that is the largest brain development study, to understand the role of biological and environmental influences on externalizing problems and behaviors. The ABCD Study has a hierarchically nested design, specifically, individuals are nested within families, which in turn are nested within a study site. In addition, the ABCD Study contains high-dimensional data from heterogeneous views or data types (e.g., brain imaging, mental health \& substance use screenings, physical health measurements) on up to 90K features, excluding SNP data, on each of the 11,878 subjects \citep{saragosa-harris_practical_2022}. The ABCD Study has been used to understand diverse phenomena, including Early Life Adversity (ELA): The experience of negative events early in life that can impact a youth's developmental trajectory \citep{orendain_data-driven_2023, brieant_characterizing_2023}. 
High-dimensional heterogeneous views make manual identification of relevant features labor-intensive and prone to differential curation. This is especially true for analyses of ELA since its broad definition means that it can be defined by a large collection of variables \citep{orendain_data-driven_2023, brieant_characterizing_2023}. In this paper we will integrate multiple brain imaging data types with ELA measures and identify the relevant features predictive of externalizing problems. 
To this end, we develop a Bayesian mixed modeling framework (BIPmixed) that accounts for the ABCD Study's nested sampling design while performing feature selection and outcome modeling.



BIPmixed is a form of Bayesian multi-view learning, where multiple views are integrated into one analysis. A naive form of multi-view learning is to concatenate views and use the larger matrix to predict some outcome vector $\bm{y}$. An alternative, though also naive, approach is to fit predictive models from each view to the outcome $\bm{y}$ and in a second step combine the resulting predictions $\hat{\bm{y}}$ by taking the mean for example. The former is ``early fusion'' and the latter ``late fusion''. These approaches are naive in that they do not account for the underlying structure and relationships between views in modeling the outcome. In contrast to early and late fusion, ``cooperative learning'' aims to simultaneously estimate shared latent structure in views and model the outcome of interest, and BIPmixed can be considered a cooperative learning method \citep{ding_cooperative_2022}. 2-step multi-view learning algorithms are an alternative to the aforementioned approaches. These methods primarily focus in the first step on dimension reduction of multiple views into only shared (i.e., joint) structure, e.g., \cite{argelaguet_multiomics_2018, argelaguet_mofa_2020}, or both shared and view-specific structure, e.g., \cite{lock_joint_2013, shen_integrative_2009, shen_sparse_2013}. This can be thought of as an integrative or multi-view factor analysis. In the second step, outputs from multi-view factor analyses are usable in an outcome model. This approach has similar issues to early and late fusion in that the outcome $\bm{y}$ is not included in latent structure estimation and feature selection, so results from this approach cannot be interpreted in direct connection to an outcome. Thus, we  focus on cooperative learning.

Cooperative learning methods have been developed to simultaneously integrate multiple views and, typically, associate a dimension reduced form of these views with an outcome $\bm{y}$. For example, canonical correlation analysis-based approaches have been proposed but are limited to at most 2 views \citep{luo_canonical_2016}. DIABLO, another multi-view method, can integrate more than 2 views but can only be used for discrete $\bm{y}$ classification \citep{singh_diablo_2019}. Frequentist multi-view methods have been developed for the integration of 2 or more views, while allowing for classification or continuous outcome prediction: JACA \citep{zhang_joint_2018}, deep IDA \citep{wang_deep_2024} and SIDA \citep{safo_sparse_2022}. Several Bayesian integrative and outcome modeling methods are also available for joint feature selection and prediction \citep{klami_bayesian_2013, klami_group_2015, yuan_patient-specific_2011, wang_ibag_2013, chekouo_bayesian_2017, mo_fully_2018, chekouo_bayesian_2021}; however, these methods generally do not account for the hierarchically nested data typical of large cohort studies. If observations from the same node of a hierarchy (e.g., subjects from a family in the same research site) are correlated but are treated as independent units, we limit our inferential capacity.

We extend Bayesian Integrative analysis and Prediction (BIP) \citep{chekouo_bayesian_2021}, which is a cooperative learning framework that allows for simultaneous view integration and outcome modeling. Furthermore, BIP allows for feature selection that is interpretable and incorporation of prior knowledge. Our extension, BIPmixed, builds on the strengths of BIP but also accounts for the hierarchically nested data that characterizes the ABCD Study and is typical of large cohort studies. Also, in BIPmixed, (clinical) covariates (e.g., sex, age, BMI, etc.) are included in the outcome model, rather than as a separate view, which can be useful when covariates are strongly predictive of the outcome \citep{cheng_wide_2016}. In Section \ref{sec-meth}, we describe BIP and our proposed extension, BIPmixed. In Section \ref{sec-results}, we show BIPmixed's performance, using simulations and the analysis of data from the ABCD Study.
We discuss strengths and limitations in Section \ref{sec-discussion}. 

\hypertarget{sec-meth}{%
\section{Methods}\label{sec-meth}}

\subsection{Multi-view Factor Analysis Framework}\label{FactorAnalysis}
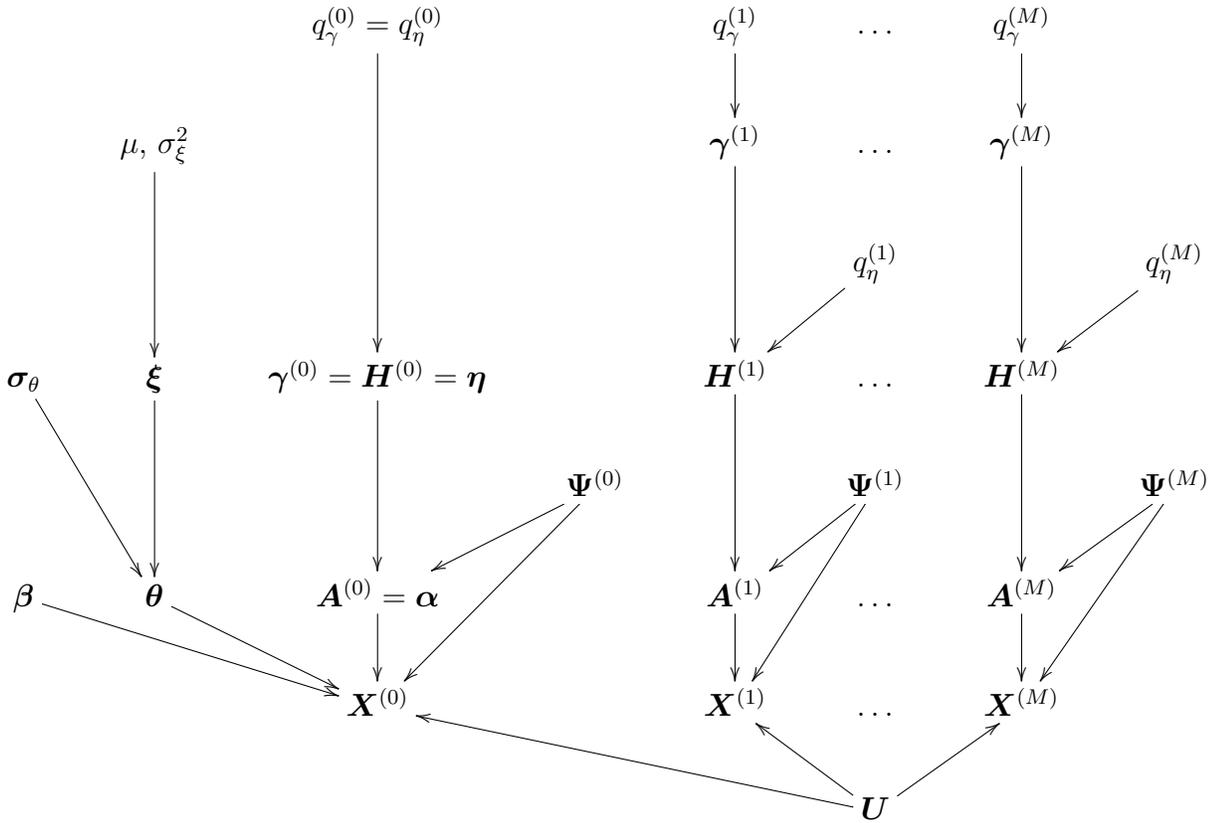
\begin{figure}
\caption{Graphical representation of the proposed probabilistic model that connects the $M+1$ views. This method integrates $M+1$ data views comprised of the outcome of interest $y$ and other views (e.g. imaging, or multi-omics data). The correlation between different views is modeled through the shared matrix $\bm{U}$, and individual sources of variation are modeled through the individual loading matrices $\bm{A}^{(m)}$ for $m \in \{0, 1, \dots, M \}$. Binary vectors and matrices $\bm{\gamma}^{(m)}$, $\bm{H}^{(m)}$ are incorporated for feature and latent component selection. In addition to the feature/component selection indicators, the outcome view $\bm{X}^{(0)}$ also incorporates fixed effects $\bm{\beta}$ as well as hierarchically nested random effects $\bm{\theta}$ and $\bm{\xi}$}
\label{graphmodel}
\[
\xymatrix{
& & q_{\gamma}^{(0)} = q_{\eta}^{(0)} \ar[ddd] & & q_{\gamma}^{(1)} \ar[d] & \dots & q_{\gamma}^{(M)} \ar[d] \\
 & \mu \text{, } \sigma_\xi^2 \ar[dd] & & & \bm{\gamma}^{(1)} \ar[dd] & \dots & \bm{\gamma}^{(M)} \ar[dd] \\
& & & & & q_{\eta}^{(1)} \ar[dl] & & q_{\eta}^{(M)} \ar[dl] \\
\bm{\sigma}_{\theta} \ar[ddr] & \bm{\xi} \ar[dd] & \bm{\gamma}^{(0)} = \bm{H}^{(0)} = \bm{\eta} \ar[dd] & & \bm{H}^{(1)} \ar[dd] & \dots &  \bm{H}^{(M)} \ar[dd] \\
& & & \bm{\Psi}^{(0)} \ar[dl] \ar[ddl] & & \bm{\Psi}^{(1)} \ar[dl] \ar[ddl] & & \bm{\Psi}^{(M)} \ar[dl] \ar[ddl] \\
\bm{\beta} \ar[drr] & \bm{\theta} \ar[dr] & \bm{A}^{(0)} = \bm{\alpha} \ar[d] & & \bm{A}^{(1)} \ar[d] & \dots & \bm{A}^{(M)} \ar[d] &\\
& & \bm{X}^{(0)} & & \bm{X}^{(1)} &\dots & \bm{X}^{(M)} & \\
& & & & & \bm{U} \ar[ulll] \ar[ul] \ar[ur]  &  &  & 
}
\]
\end{figure}

In Figure \ref{graphmodel}, we describe the Bayesian factor analysis framework in which observed data are divided into \( M+1 \) views. We integrate information across views  $\bm{X}^{(m)} \), \( m = 0, 1, 2, \dots, M \), by decomposition into a latent factor \( \bm{U} \), view-specific loadings \( \bm{A}^{(m)} \), and Gaussian noise. Each view is \(\bm{X}^{(m)} = \bm{U A}^{(m)} + \bm{E}^{(m)},\) where \( \bm{X}^{(m)} \in \mathbb{R}^{n \times p_m} \) is the observed matrix for view \( m \), \( n \) is the number of observations, and \( p_m \) is the number of features in data type \( m \). \( \bm{U} \in \mathbb{R}^{n \times r} \) is the factor matrix shared across views, and \( \bm{U} \sim \mathcal{N}(0, \bm{I}_{nr}) \) where \( \bm{I}_{nr} \) is the identity matrix and $r$ is the number of components the latent factor consists of. We use $l$ to denote the latent factor component index, \( l=1, \dots, r \). \( \bm{A}^{(m)} = (a_{lj}^{(m)}) \in \mathbb{R}^{r \times p_m} \) encodes loadings specific to each view $m$. \( \bm{E}^{(m)} \) is a noise matrix, specific to each view, that is normally distributed \( \bm{E}^{(m)} \sim \mathcal{N}(\bm{0}, \bm{\Psi}^{(m)}) \). \( \bm{\Psi}^{(m)} = Diag \left(\sigma_1^{2(m)}, \sigma_2^{2(m)}, \dots, \sigma_{p_m}^{2(m)} \right) \otimes I_n\) where \( \sigma_j^{2(m)} \) is the variance for the $j$th feature in the $m$th view. Outcome \( \bm{y} \) is incorporated as a view, \( \bm{y} = \bm{X}^{(0)}\), after being adjusted for covariates and random effects.

\subsection{Stochastic Search Feature and Latent Component Selection}\label{featureselection}
We aim to identify a set of features that are associated across views and  biologically interpretable. As in BIP \citep{chekouo_bayesian_2021}, we introduce then binary indicator variables: \( \eta_{lj}^{(m)} \) is a binary feature selection indicator, and \( \gamma_l^{(m)} \) is a latent factor component selection indicator. For outcome view $\bm{X}^{(0)}$,  $\eta_{l1}^{(0)}={\gamma}_l^{(0)}$. Prior for \( \eta_{lj}^{(m)} \) is a mixture between a point mass at 0, $\delta_0$ (indicating feature $j$ excluded in $l$th latent component) and Bernoulli \( q_\eta \), and \( \gamma_{l}^{(m)} \) is Bernoulli \( q_\gamma \):
    $\eta_{lj}^{(m)} | \gamma_l^{(m)} \sim \left(1 - \gamma_l^{(m)}\right) \delta_0 + \gamma_l^{(m)} \text{Bernoulli}(q_\eta)$.
Factor loading prior for \( a_{lj}^{(m)} \) depends on both \( \gamma_l^{(m)} \) and \( \eta_{lj}^{(m)} \). The prior is a mixture of $\delta_0$ and a normal distribution where $\tau_{lj}^2$ is a shrinkage parameter:
   $a_{lj}^{(m)} | \gamma_l^{(m)}, \eta_{lj}^{(m)} \sim \left(1 - \gamma_l^{(m)} \eta_{lj}^{(m)}\right) \delta_0 + \gamma_l^{(m)} \eta_{lj}^{(m)} \mathcal{N} \left( 0, \tau_{lj}^2 \sigma_j^{2(m)} \right)$.  
If, for example, the $l$th latent component indicator $\gamma_l^{(m)}$ or the indicator for the $j$th feature in the $l$th component $\eta_{lj}^{(m)}$ is equal to 0, then the loading for the $j$th feature in the $l$th component is set to $0$ with probability 1. Otherwise, we have normal loadings that have variance dependent on $\tau_{lj}^2$ and $\sigma_j^{2(m)}$.

\subsection{Mixed Outcome Model}\label{OutcomeModel}
In the outcome model, fixed effects \( \bm{\beta} \) correspond to the design matrix \( \bm{W} \), and the random effects \( \bm{\theta} \) are included by the design matrix \( \bm{Z} \). At the individual observation level, for individual $i$ in family $f$ within site $s$, 
the mixed outcome model is defined as 
    $$y_i = y_{ifs} = \bm{W}_{i} \bm{\beta} + \bm{Z}_{i}\bm{\theta} + \bm{U}_{i}\bm{\alpha} + \varepsilon_{i} = \bm{W}_{i} \bm{\beta} + \theta_{f:s} + \bm{U}_{i}\bm{\alpha} + \varepsilon_{i},$$

where \( \bm{U}_{i} \) is the $i$th row of the latent factor.
We drop feature index $j=1$ and $m=0$ in the outcome view from the following notation. The outcome view's loadings \( \bm{\alpha} \) follow a spike and slab prior distribution: \( a_{l} \mid \gamma_l \sim \left( 1-\gamma_l \right)\delta_0 + \gamma_l N\left( 0, \tau_l^2 \sigma^2 \right) \). \( \bm{W}_{i} \) is the $i$th row from the fixed effect design matrix of dimension $p_\beta$, the number of covariates. \( \bm{\beta} \) is the fixed effects coefficient vector of length $p_\beta$, \( \theta_{f:s} \) is the contribution of family \( f \) in site \( s \) to the intercept, and \( \varepsilon_{i} = \varepsilon_{ifs} | \sigma^2 \overset{iid}{\sim} \mathcal{N}(0, \sigma^2) \). Priors for the fixed and random effects are specified as follows: $\bm{\beta} \sim \mathcal{N}(0, \sigma_{\bm{\beta}}^2\bm{I}_{p_\beta})$ ; $\mu \sim \mathcal{N}(0, \sigma_{\mu}^2)$; $\xi_s \mid \mu, \sigma_\xi^2 \overset{iid}{\sim} \mathcal{N}(\mu, \sigma_{\xi}^2)$ ; $\theta_{f:s} \mid \xi_s, \sigma_{\theta_s}^2 \overset{iid}{\sim} \mathcal{N}(\xi_s, \sigma_{\theta_s}^2)$ ; $\sigma_{\xi}^2 \sim \text{IG}(a_{\xi}, b_{\xi})$ ; $\sigma_{\theta_s}^2 \overset{iid}{\sim} \text{IG}(a_{\theta}, b_{\theta})$ ; $\sigma^2 \sim \text{IG}(a_{\sigma}, b_{\sigma})$
where $IG(a,b)$ is inverse gamma with the shape parameter $a$ and scale $b$. We can interpret $\mu$ as the grand mean or intercept and $\xi_s$ as the site $s$ effect, which centers the $\theta_{f:s}$ family in site effects. Variance parameters $\sigma_\beta^2$, $\sigma_\mu^2$, $\sigma_\xi^2$, and $\sigma_{\theta_s}^2$ represent the fixed effect variability, that in the grand mean, the site effects, and family in site effects respectively. Notationally, we let $\bm{\sigma_\theta^2} = \left( \sigma_{\theta_1}^2, \dots, \sigma_{\theta_s}^2 \right)$ as the vector of site-specific variances. Section \ref{Hyperparameters} details hyperparameter specification. 

\subsection{Posterior Inference}
By Markov Chain Monte Carlo (MCMC) sampling, we jointly estimate feature selection indicators \( \bm{H}^{(m)} = (\eta_{lj}^{(m)})_{l \in \{1, \dots, r \}, j \in \{1, \dots, p_m \}} \), latent factor indicators \( \bm{\gamma}^{(m)} = (\gamma_{l}^{(m)})_{l \in \{1, \dots, r \}} \), shared latent factor \( \bm{U} \), loadings \( \bm{A}^{(m)} \), fixed and random effects \( \bm{\beta} \), \( \bm{\xi} \) and \( \bm{\theta} \). We sample all feature and component indicators \( \eta_{lj} \) and \( \gamma_l \) using Metropolis-Hasting steps, after integrating the loadings \( \bm{A}_{(\eta)}^{(m)} \) out of the data model \( \bm{X}^{(m)} = \bm{U}\bm{A}_{(\eta)}^{(m)} + \bm{E}^{(m)} \).  Then, all other parameters are sampled by Gibbs steps. Note, subscript $(\eta)$ in $\bm{A}_{(\eta)}^{(m)}$ indicates the submatrix of $\bm{A}$ where $\eta_{lj}=1$. 

\begin{enumerate}
    \item Initialize $\sigma_{j}^{2(m)} = 1$ for all $m \in \{0, 1, \dots, M \}$ and $j \in \{1, \dots, p_m \}$, $\sigma_{\xi}^{2} = 1$, $\sigma_{\theta_s}^{2} = 0.5$ for all $s \in \{1, \dots, N_S \}$, $\mu = \bar{\tilde{y}}$, $\bm{\beta} = (\bm{W}^T \bm{W})^{-1}\bm{W\tilde{y}}$, and the rest from their priors,
    \item After integrating out the loadings \( \bm{A}_{(\eta)}^{(m)} \) from the data model \( \bm{X}^{(m)} = \bm{U}\bm{A}_{(\eta)}^{(m)} + \bm{E}^{(m)} \), a Metropolis-Hastings step is used for each of latent component and feature selection indicators \( \bm{\gamma}, \bm{H} \), as described \cite{chekouo_bayesian_2021}'s supplementary materials, 
    \item Gibbs steps are used to sample \( \bm{A}_{(\eta)}^{(m)} \), \( \sigma_j^{2(m)} \) , \( \bm{U} \) from their full conditionals in \cite{chekouo_bayesian_2021},
    \item Residualize outcome view on current iteration's \( \bm{U}\bm{\alpha} \), and Gibbs step outcome model parameters, in order, $\bm{\beta}$ (if covariates are available), $\bm{\theta}$, $\bm{\xi}$, $\bm{\sigma_\theta^2}$, $\sigma_\xi^2$, $\mu$. Section \ref{OutcomeSamplingDetails} below are details about sampling parameters specific to the outcome view. 
\end{enumerate}
\subsubsection{Outcome Model-Specific Parameter Sampling}
\label{OutcomeSamplingDetails}
Since the site-level random effects follow 
$\bm{\xi}|\mu,\sigma_\xi^2 \sim \mathcal{N}\left(\mu \bm{1}, \sigma_\xi^2\bm{I}_{N_S} \right)$, then the full conditional distribution of the grand mean $\mu$ is as  $ \mu|\bm{\xi},\sigma_\xi^2 \sim \mathcal{N}\left(m_{\mu}, \Sigma_{\mu}\right)$, where
$\Sigma_{\mu} = \left(1/\sigma_{\mu}^2 + N_S/\sigma_{\xi}^2\right)^{-1}$ and  $m_{\mu} = \Sigma_{\mu} (\sum_{s=1}^{N_S}\xi_s/\sigma_{\xi}^2)$. 
The full conditional distribution of the fixed effect vector $\bm{\beta}$ is a multivariate normal distribution $\boldsymbol{\beta} | \text{rest} \sim \mathcal{N}(\mathbf{\mu}_{\beta}, \mathbf{\Sigma}_{\beta})$ where  $\mathbf{\Sigma}_{\beta} = \left(\sigma_{\beta}^{-2}\mathbf{W}^\top \mathbf{W} + \sigma^{-2}\bm{I}_n\right)^{-1}$, $\mathbf{\mu}_{\beta} = \mathbf{\Sigma}_{\beta} \left( \mathbf{W}^\top (\bm{y} - \bm{Z}\bm{\theta} - \bm{U \alpha}) /{\sigma^2}  \right)$, and "rest" indicates conditioning on all other parameters. For site-level random effect vector $\bm{\xi} = (\xi_1, \dots, \xi_{N_S})^T$, for each site $s$, from families in that site, the full conditional of $\xi_s$ is   $\xi_s | \text{rest} \sim \mathcal{N} \left(\mu_{\xi_s}, V_{\xi_s}\right)$ with $\mu_{\xi_s} = V_{\xi_s}\left(\frac{1}{\sigma_{\theta_s}^2} \sum_{f \in \mathcal{F}_s} \theta_{f:s}\right)$ and $V_{\xi_s} = \left(\frac{1}{\sigma_{\xi}^2} + \frac{n_{s}}{\sigma_{\theta_s}^2}\right)^{-1}$ where $\mathcal{F}_s$ is the set of families in site. The full conditional distribution of the family-level random effects $\bm{\theta}$ is also a multivariate normal distribution $\theta_{f:s} | \text{rest} \sim \mathcal{N}(\mathbf{\mu}_{\theta_{f:s}}, \mathbf{\Sigma}_{\theta_{f:s}})$ with $\Sigma_{\theta_{f:s}} = \left( \frac{1}{\sigma_{\theta_s}^2} + \frac{n_{f:s}}{\sigma^2} \right)^{-1}$ and $\mu_{\theta_{f:s}} = \Sigma_{\theta_{f:s}} \left( \frac{\xi_s}{\sigma_{\theta_s}^2} + \frac{n_{f:s}}{\sigma^2} \bar{\tilde{y}}_{sf+} \right)$ where $n_{f:s}$ is the number of observations in family $f$ at site $s$, and $\bar{\tilde{y}}_{sf+}$ is the sample mean of the residualized outcome $\tilde{\bm{y}} = \bm{y}  - \bm{W\beta} - \bm{U\alpha}$ over the family $f$ at site $s$. Full conditionals of site-level Variance ($\sigma_{\xi}^2$) and site-specific family-level variances  ($\sigma_{\theta_{s}}^2$) are defined as: $\sigma_\xi^2 | \text{rest} \sim \text{IG} \left( a_\xi + \frac{N_S}{2}, \; b_\xi + \frac{1}{2} \sum_{s=1}^{N_s} \xi_s^2 \right)$ and $\sigma_{\theta_s}^2 | \text{rest} \sim \text{IG} \left( a_\theta + \frac{n_{s}}{2}, \;b_\theta + \frac{1}{2} \sum_{f \in \mathcal{F}_s} (\theta_{f:s} - \xi_s)^2 \right)$ with \( N_S \) number of study sites and \( n_{s} \) number of families in site \( s \). The full conditional of the outcome error
 variance $\sigma^2$ is $\sigma^2 | \text{rest} \sim \text{IG} \left( a_{\sigma} + \frac{N}{2}, b_{\sigma} + \frac{1}{2} \bm{\tilde{y}}^{T}\Sigma_\sigma^{-1}\bm{\tilde{y}} \right)$ where $\tilde{\bm{y}} = \bm{y} - \bm{W\beta} - \bm{Z\theta}$ and $\Sigma_\sigma^{-1} = \left( \bm{U}_{(\gamma=1)}\bm{U}_{(\gamma=1)}^T + \bm{I}_n \right)^{-1}$. 

\subsection{Prediction}\label{Prediction}

We predict for a new set of individuals that belong to a known site $s$ with $\bm{X}_{\text{new}}^{(m)}$, $m = 1, \dots, M$ observed. Given a model $\{ \bm{\gamma}^{(m)},\bm{H}^{(m)}: m=0,1,\dots,M \}$ which consists of the set of latent component selection indicator vectors and feature selection matrices, respectively, we estimate  loadings as $
    \hat{\boldsymbol{a}}_{. j(\eta)}^{(m)}=\hat{\sigma}_j^{2(m)}\left(\overline{\boldsymbol{U}}_{(\gamma)}^T \overline{\boldsymbol{U}}_{(\gamma)} + I_{n_\gamma} \right)^{-1} \overline{\boldsymbol{U}}_{(\gamma)}^T \boldsymbol{x}_{.j}^{(m)}$,    
where posterior mean estimates $\hat{\sigma}_j^{2(m)}$ and $\overline{\boldsymbol{U}}$ are used, and subscripts $(\gamma)$ and $(\eta)$ indicate elements were $\gamma_l=1$ or $\eta_{lj}=1$. For the outcome model loadings, we let $\boldsymbol{x}_{.j}^{(0)}=\boldsymbol{y}-\boldsymbol{W}\hat{\boldsymbol{\beta}}-\boldsymbol{Z}\hat{\boldsymbol{\theta}}$ where $\hat{\boldsymbol{\beta}}$ and $\hat{\boldsymbol{\theta}}$ are  posterior mean estimates for obtaining $\hat{\boldsymbol{a}}_{. j(\eta)}^{(0)}=\hat{\boldsymbol{\alpha}}_{(\gamma)}$. We use $\hat{\boldsymbol{\xi}}$ as an estimate for $\hat{\boldsymbol{\theta}}$ since the family effect in the test set has not been observed. From $\hat{\boldsymbol{A}}_{(\eta)}=\left( \hat{\boldsymbol{a}}_{. j(\eta)}^{(1)}, \dots, \hat{\boldsymbol{a}}_{. j(\eta)}^{(M)} \right)$ and $\boldsymbol{X}_{\text{new}}=\left( \boldsymbol{X}^{(1)}_{\text{new}}, \boldsymbol{X}^{(2)}_{\text{new}}, \dots, \boldsymbol{X}^{(M)}_{\text{new}} \right)$, both column-wise concatenated matrices, and $\bm{D(\hat{\sigma}}^{-2})$, a diagonal matrix with posterior means ${\hat{\sigma_j}^{2(m)}, m=1, \dots, M, j=1, \dots, p_m}$ as elements, we estimate $
    \hat{\boldsymbol{U}}_{\text{new},(\gamma)}=\left( \hat{\boldsymbol{A}}_{(\eta)} \bm{D(\hat{\sigma}}^{-2}) \hat{\boldsymbol{A}}_{(\eta)}^T + I_r\right)^{-1} \hat{\boldsymbol{A}}_{(\eta)} \bm{D(\hat{\sigma}}^{-2}) \boldsymbol{X}_{\text{new}}$. With $\boldsymbol{W}_{\text{new}}$ and $\boldsymbol{Z}_{\text{new}}$ for new individuals' fixed and random effect design matrices, we have, for 1 model, $
    \hat{\boldsymbol{y}}_{\text{new}} = \hat{\boldsymbol{U}}_{\text{new},(\gamma)}\boldsymbol{\alpha}_{(\gamma)} + \boldsymbol{W}_{\text{new}} \hat{\bm{\beta}} + \boldsymbol{Z}_{\text{new}} \hat{\bm{\theta}}$.
From Bayesian model averaging over a maximum number of models, which are selected as the models with largest posterior probability, the prediction of new subjects is estimated as: $
    \hat{\boldsymbol{y}}_{\text{new}} = \boldsymbol{W}_{\text{new}} \hat{\bm{\beta}}  + \boldsymbol{Z}_{\text{new}} \hat{\bm{\theta}} + \sum_{\boldsymbol{\eta},\boldsymbol{\gamma}} \hat{\boldsymbol{U}}_{\text{new},(\gamma)}\boldsymbol{\alpha}_{(\gamma)} p \left( \boldsymbol{\gamma}, \boldsymbol{\eta} | \overline{\boldsymbol{U}}_{(\gamma)}, \boldsymbol{X} \right)$.

\subsection{Hyperparameter Specification}\label{Hyperparameters}

We choose minimally a latent component number $r$ equal to the number of views, including the outcome view, $M+1$, to be analyzed such that every component can be associated with only one view. For example, if gene expression and SNP data are included as 2 views in modeling outcome $y$, let $r_{\text{min}}=3$. To get a preliminary understanding of the data's latent structure, we concatenate the non-outcome views and $\bm{y}$ column-wise, standardize, and then calculate the sample covariance matrix and its associated eigenvalues. After scree plot inspection, we choose $r$ where eigenvalues stabilize. Additionally, the loadings shrinkage parameter $\tau_{lj}^2$ is fixed to 1. Its properties are explored more in \cite{chekouo_bayesian_2017}. The maximum number of models in Bayesian Model Averaging (BMA) prediction is 50, since more models give similar results. Non-outcome view feature selection prior parameter $q_\eta$ set to 0.05 by the assumption of sparse feature importance. Fixed effect vector $\bm{\beta}$ has an uninformative normal prior with uncorrelated fixed effects, the same variance $\sigma_\beta^2$ fixed at a large value of 100 and prior mean 0. Other priors' hyperparameters are set uninformatively, and particular values are available as default arguments in the BIPmixed implementation available from Section \ref{code-availability}. 

\subsection{ABCD Study Data Processing}
\label{ABCDDataProcessing}
The ABCD Study is the largest study of brain development. We performed a cross-sectional analysis using baseline (n=11,878) from the 5.1 data release. To study Early Life Adversity (ELA), we used the union of features from \citet{brieant_characterizing_2023} and \citet{orendain_data-driven_2023}. We filtered features similar to \citet{brieant_characterizing_2023} and \citet{orendain_data-driven_2023} by excluding ELA features with $>50\% $ missingness and $< 0.05 \%$ endorsement. Endorsement filtering resulted in dropping 1 feature from Orendain et al. 2023. 
These decisions led to an ELA view of 88 features. We included structural MRI (sMRI) metrics as 2 different views: cortical surface area (sMRI\_SA) and cortical thickness (sMRI\_CT), parcellated by the Destrieux Atlas, which has 74 regions of interest (ROIs). With 74 ROIs across 2 hemispheres, we had 148 features in sMRI\_CT and sMRI\_SA respectively. Resting-state functional MRI (fMRI) connectivity between ROIs from the Gordon Network is the 4th view. Connectivity measures included pairwise and self-correlations between 13 networks, resulting in $13^2=169$ features. Outcome $y$ is Externalizing Problems (R-Score) obtained from the Child Behavior Checklist \citep{achenbach_child_2000}. 

\begin{figure}
    \centering
    \caption{Flowchart of sample selection from the ABCD Study at baseline (n=11,878). Subjects excluded for missing any data in each view, Early Life Adversity (ELA), structural MRI Surface Area (sMRI SA), structural MRI Cortical Thickness (sMRI CT), functional MRI correlations between Gordon Atlas Regions of Interest (fMRI Corr), covariates, and outcomes feature Externalizing Problems. Resulting sample size n=7,370. On stratifying by study site, data is repeatedly split into 20 train and test sets with 80:20 ratio respectively, which results in train $N_{avg}=5885.5$ and test $N_{avg}=1484.5$.}
    \includegraphics[width=1\linewidth]{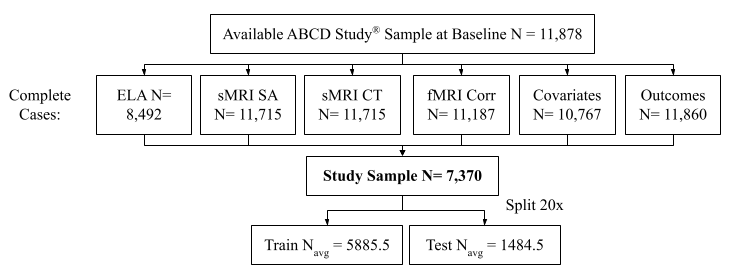}
    \label{fig:FlowChart}
\end{figure}

We describe the covariates of the selected sample in Table \ref{SampleDescription}, and explore the impact of covariate inclusion on prediction by performing the analysis with and without covariates included in the modeling and subsequently evaluate the Mean Square Prediction Error (MSE).

\begin{table}[ht]
\centering

\caption{Externalizing (R-scores) mean (SD) 4.03 (5.44), sex, baseline age, parent marital status, and race/ethnicity as Native American, Asian, Black, Hispanic/Latinx, Pacific Islander, White, or Other Race Count (\%), which can add to more than $100\%$ since endorsement of more than 1 is allowed. Ordinal categorical features family income in the past 12 months 7.44 (2.31) and parental highest education 17.37 (2.47) are treated as continuous. }

\begin{tabular}{lll}
\hline
\textbf{Variable} & \textbf{Level} & \textbf{Value} \\
\hline
n &  & 7370 \\
Externalizing Problems (Raw) (mean (SD)) &  & 4.03 (5.44) \\
Sex (At Birth) (\%) & Male & 3861 (52.4) \\
                    & Female & 3509 (47.6) \\
Age (Months) (mean (SD)) &  & 119.07 (7.49) \\ 
Race/Ethnicity Count (\%) & Native American  
& 5860 (79.5) \\
& Asian 
& 1247 (16.9) \\
& Black 
& 225 (3.1) \\
& Hispanic/Latinx 
& 47 (0.6) \\
& Pacific Islander 
& 475 (6.4) \\
& Other Race 
& 439 (6.0) \\
& White 
& 1364 (18.5) \\ 
Total Family Income (Past 12 Months) (mean (SD)) &  & 7.44 (2.31) \\
Highest Parent Education Completed (mean (SD)) &  & 17.37 (2.47) \\ 
Parent Marital Status (\%) & Married & 5446 (73.9) \\
                           & Widowed & 43 (0.6) \\
                           & Divorced & 576 (7.8) \\
                           & Separated & 217 (2.9) \\
                           & Never married & 708 (9.6) \\
                           & Living with partner & 380 (5.2) \\
\end{tabular}
\label{SampleDescription}
\end{table}

Subjects are excluded for missing data in any view, covariates, and Externalizing Problems, leading to n=7,370 (Figure \ref{fig:FlowChart}). The ABCD Study is hierarchically nested in that subjects are nested in families in 22 study sites across the United States, including, for example, Children’s Hospital Los Angeles, Florida International University, and the Laureate Institute for Brain Research (LIBR). 

\begin{figure}
    \centering
    \caption{Scree plot of eigenvalues associated with column-wise concatenated ABCD Study train views’ covariance matrix latent components, including Externalizing Behaviors (R-Score) as the outcome. The plots suggest $r=6$ is a sufficient number of latent components to include in each model, as eigenvalues beyond the 6th component stabilize.}
    \includegraphics[width=1.0\linewidth]{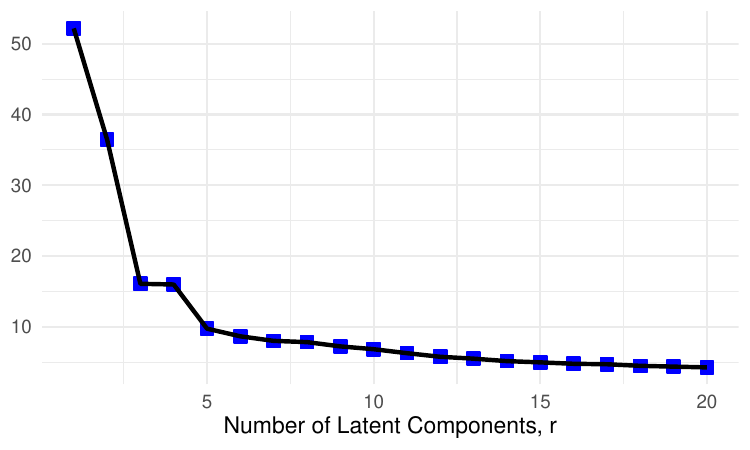}
    \label{fig:sensitivity-plot}
\end{figure}

Finally, for the data analysis, to choose $r$ - the number of latent factors - we compute the covariance matrix by first concatenating the 4 views, the outcome externalizing problems, and the covariates column-wise. We standardize the resulting data and calculate the covariance matrix, and its associated eigenvalues to choose hyperparameter $r=6$ where the eigenvalues level off (Figure \ref{fig:sensitivity-plot}).

\subsection{Competing Methods}

We compare BIPmixed against the original BIP framework, which does not account for random effects, in both the ABCD Study data analysis and the simulation studies \citep{chekouo_bayesian_2021}. We also evaluate performance in simulations against Cooperative Learning Lasso with hyper-parameter $\rho=0.5$, i.e. halfway between early and late fusion, using 10-fold cross-validation (CV) on the training set to choose the penalty parameter \citep{ding_cooperative_2022}. Furthermore, we include a 2-step method: PCA2Step, which does not account for multiview data structure. PCA2Step starts with Principal Components Analysis of concatenated train set views, and then, in the second step, the top $r=4$ principal components are used in a frequentist linear mixed model. 

\subsection{Simulation Scenarios}\label{SimulationScenarios}

For each scenario described in the proceeding section, we simulate $S=20$ training and test datasets. We use a method for generating views used in high-dimensional data integration studies \citep{luo_canonical_2016, chekouo_bayesian_2021}. We assume 4 views $\boldsymbol{X}=\left(\boldsymbol{X}^{(1)}, \boldsymbol{X}^{(2)}, \boldsymbol{X}^{(3)}, \boldsymbol{X}^{(4)}\right)$ where $\boldsymbol{X}^{(m)} \in \Re^{n=800 \times p=500}$. In each $\boldsymbol{X}^{(m)}$, the first 100 features form groups where there are 10 main features each connected to 9 supporting features. As a consequence, in each view, there are $p-100$ singletons. Intra-view correlation is \(\Psi^{(m)} = \begin{pmatrix} \overline{\Psi}_{100 \times 100} & 0 \\ 0 & I_{p - 100} \end{pmatrix}\). $\overline{\Psi}_{100 \times 100}$ is block diagonal with block size 10, between-block correlation 0, within block $9 \times 9$ compound symmetry in the supporting features with elements equal to $0.7^2$, and correlation between a main feature and a supporting feature is $0.7$. We assume feature groups contribute to correlation between views by the loadings: $\boldsymbol{A}=\left(\boldsymbol{A}^{(1)}, \boldsymbol{A}^{(2)}, \boldsymbol{A}^{(3)}, \boldsymbol{A}^{(4)}\right)$. The first 100 columns $\boldsymbol{A}^{(m)} \in \Re^{r=4 \times p}$ corresponding to the loading for the first 100 features aforementioned are sampled from independent and identically distributed (i.i.d.) uniform distribution on $\left[ -0.5, -0.3 \right] \cup \left[ 0.3, 0.5 \right]$, and the main features' loadings are multiplied by 2. The remaining $p-100$ features are set to 0. Each element of latent factor $\boldsymbol{U} \in \Re^{n \times r}$ is generated from i.i.d. standard normal. The outcome $y$ in view $m=0$ is generated with $\boldsymbol{A}^{(0)}=\alpha=\left( 1,1,1,0 \right)$, i.e., 3 latent components predictive of $y$. For both the train and test sets, we set $n=800$ by letting the number of study sites $n_s=20$, the number of families per site $n_{f:s}=20$, and the number of individuals per family to $n_{i:f:s}=2$. Study sites are shared across train and test, and families are unique to train and test. Grand $\mu=1$ and random effects $\xi$ and $\theta_{f:s}$ are generated from 
    $\xi_s \mid \mu, \sigma_\xi^2 \overset{iid}{\sim} \mathcal{N}(\mu, \sigma_{\xi}^2)$ and
    $\theta_{f:s} \mid \xi_s, \sigma_{\theta_s}^2 \overset{iid}{\sim} \mathcal{N}(\xi_s, \sigma_{\theta_s}^2)$. 

For all features, including the outcome $y$ in view $m=0$, residual variance $\sigma^2=1$. Parameters $\sigma_{\xi}^2$ and $\sigma_{\theta_s}^2$ are different in each simulation scenario. We fix three scenarios describing within-site (i.e., family:site) variance \(\sigma_{\theta_s}^2\) and between-site variance \(\sigma_{\xi}^2\), where, for a given scenario, \(\sigma_{\theta_s}^2\) is fixed to a specific value across all study sites. In Scenario 1, no random effects are included, with both random effect variances set to 0. Scenario 2 assumes the within-site variance is greater than the between-site variance, with \(\sigma_{\theta_s}^2 = 1\) and \(\sigma_{\xi}^2 = 0.5\). Conversely, Scenario 3 assumes the within-site variance is smaller than the between-site variance, with \(\sigma_{\theta_s}^2 = 0.5\) and \(\sigma_{\xi}^2 = 1\). From $\boldsymbol{U}$, $\bm{\alpha}$, $\theta_{f:s}$, and $\varepsilon_i \sim N(0,\sigma^2)$, we generate the outcome $y$ from $y_i = \bm{U}_{i}\bm{\alpha} + \theta_{f:s} + \varepsilon_{i}$.

\subsection{Evaluation Criteria}

Across the $S$ datasets, we evaluate feature selection and prediction performance. Using a Marginal Posterior Probability (MPP) threshold of 0.5, we calculate the false positive rate (FPR) and false negative rate (FNR). Additionally, to evaluate feature selection in a threshold-independent way, we estimate the AUC of feature classification as important or not. AUC of feature selection probabilities is not available for methods that only support binary feature selection, e.g. Cooperative Learning. Prediction metrics are mean square error (MSE) and Var($\hat{y}$) and are computed for all the methods.

\hypertarget{sec-results}{%
\section{Results}\label{sec-results}}

\subsection{Data Analysis Results}
\label{DataAnalysisResults}

\begin{figure}
    \centering
        \caption{BIPmixed analysis of the ABCD Study dataset with outcome \( \bm{y} \) raw externalizing problems. 
        \textbf{Panel A}. View contributions to latent factor components where a contribution is defined as the number of important features, those with marginal posterior probabilities \( >0.5 \). Views: Early Life Adversity (ELA), functional MRI (fMRI) functional connectivity, and 2 from the structural MRI (sMRI) modality, Cortical Thickness (CT) and Surface Area (SA). Latent factor components 3 and 5 identified as important to the externalizing problems outcome (MPP $>0.5$) are highlighted with a red dashed box. \textbf{Panel B}. Sankey plot important feature mapping from views to latent components with a red dashed box around important components. \textbf{Panel C}. Within study site variances \( \sigma^2_{\theta_s} \) to between study site variance \( \sigma^2_\xi \) credible intervals, with the dashed line indicating within and between site variance equivalence. Posterior mean (credible interval) for outcome model residual variance \( \sigma^{2(0)} \) is 1.255 (1.187, 1.326), and for between study site variance \( \sigma^2_{\xi} \) is 0.345 (0.118, 1.283).}
    \includegraphics[width=1\linewidth]{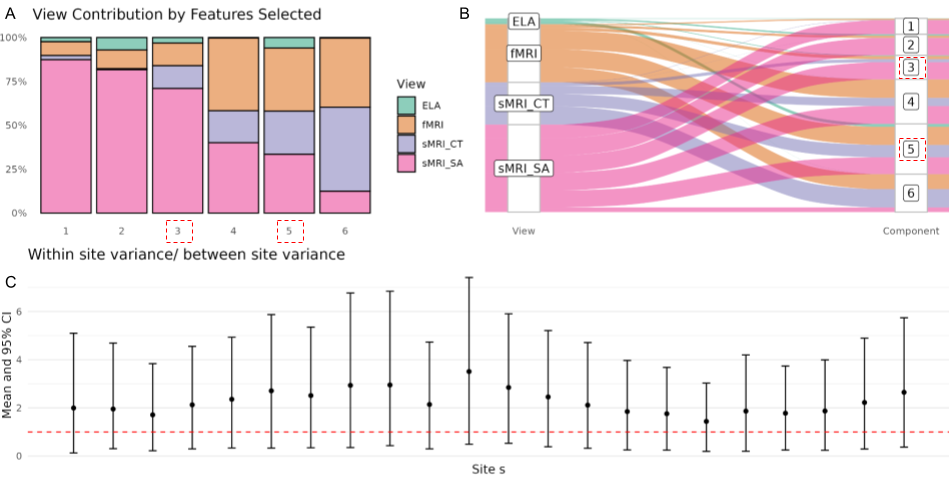}
    \label{fig:DataAnalysisResults}
\end{figure}

Figure \ref{fig:DataAnalysisResults} presents results from applying BIPmixed to the ABCD Study dataset with externalizing problems as the outcome. Each of the four views 
contributed differently across latent factor components, with contribution defined as the number of features with marginal posterior probabilities (MPPs) greater than 0.5. Figure \ref{fig:DataAnalysisResults}A shows the proportion of important features selected across different views. sMRI\_SA demonstrated notable contributions across all latent components and ELA represented the least across components. Important components associated with the outcome, as defined by a posterior mean estimate for $\gamma_l^{(0)}$ are emphasized with red dashed boxes, specifically for components $l=3,5$. Figure \ref{fig:DataAnalysisResults}B visualizes the mapping of important features from view to latent factor component using a Sankey plot, again indicating components 3 and 5 as associated with the outcome Externalizing Problems, and showing that while sMRI\_SA is represented well across all latent factor components, views sMRI\_CT and fMRI especially co-occur in components 3-6. Figure \ref{fig:DataAnalysisResults}C depicts the within and between study site variances ($\sigma^2_{\theta_s}$ and $\sigma^2_\xi$ respectively) as a forest plot with credible intervals across study sites. The dashed line in Panel C indicates where within-site and between-site variances are equivalent. While credible intervals cover the equivalence point, all posterior mean estimates for $\sigma^2_{\theta_s}$ are greater than $\sigma^2_{\xi}$, suggesting a greater within site variance than between variance in the externalizing outcome while controlling for covariates and views represented in the latent factor. Posterior estimates for residual variance and between-site variance are 1.255 (1.187, 1.326) and 0.345 (0.118, 1.283), respectively.

Latent factors 3 and 5 are considered important to externalizing problems since $\hat{\gamma}_3^{(0)}>0.5$ and $\hat{\gamma}_5^{(0)}>0.5$. We describe the top features, i.e. those with the largest MPP $\hat{\eta}$ in each component. Each component consists of features across views. For component 3, from the fMRI view, important features include functional connections from the auditory network to the following: auditory network (self-correlation), cingulo-opercular network, default network, non-Gordon network voxels, and salience network. For the ELA view, top features were measures of neighborhood safety protocols, parental awareness of the child’s location, avoidant personality problems, and reports of bullying. The sMRI\_CT view included cortical thickness in the superior precentral sulcus and middle-anterior cingulate gyrus and sulcus. Additionally, from the sMRI\_SA view, top features included cortical area in the superior occipital gyrus, and the occipital pole. For component 5, from the fMRI view, important features included a bidirectional connection between the retrosplenial temporal network and auditory network, retrosplenial temporal network (self-connectivity), salience network (self-connectivity), and a connection from the cingulo-opercular network to the ventral attention network. Top features from the ELA view were parent-reported measures of neighborhood safety protocols, self-reported neighborhood safety from crime, how often the child and parents eat dinner together, antisocial personality problems, and witnessing violence between adults in the home. The sMRI\_CT view included structural features like cortical thickness in the left hemisphere calcarine sulcus, right hemisphere cuneus, and right hemisphere occipital pole). Finally, from the sMRI\_SA view, top features were the cortical area in the left hemisphere marginal branch of the cingulate sulcus, the left hemisphere medial occipito-temporal sulcus and lingual sulcus, and the left hemisphere orbital sulci. 

In analyzing the impact of covariate inclusion in the Bayesian Integrative Mixed Model (BIP) and its mixed variant (BIPmixed) on prediction, we observed a slightly improved performance for BIPmixed when covariates were incorporated directly into the outcome model compared to BIP, where covariates were included as a separate view. Specifically, for the outcome Externalizing Problems R-Score, the Mean (SD) of MSE with covariates was 26.3 (2.66) for BIPmixed compared to 27.4 (3.01) for BIP. Without covariates, MSE was comparable between the methods, with BIPmixed yielding 27.8 (3.36) and BIP yielding 27.8 (2.41). The inclusion of covariates in BIPmixed also appeared to reduce variability in prediction performance, as evidenced by lower standard deviations.

\subsection{Simulation Results}\label{SimulationResults}

Results include the performance of 4 methods: BIP, BIPmixed, Cooperative Learning, and PCA2Step, a non-integrative comparator, across 3 scenarios that vary in random effects. Prediction performance metrics include Mean Square Error (MSE) and variance of predictions ($\text{Var}(\hat{y})$), and feature selection performance metrics are False Positive Rate (FPR), False Negative Rate (FNR), and Area Under the Curve (AUC). 

In Scenario 1, BIP achieves the lowest MSE (1.760), closely followed by BIPmixed (1.816) and Cooperative Learning (1.826). PCA2Step (2.054) performs comparably but with a higher variance ($\text{Var}(\hat{y})$ = 4.091). Cooperative Learning shows the lowest prediction variance (1.511), suggesting more precise predictions. AUC values for BIP and BIPmixed remain at 1.000, demonstrating perfect feature discriminative ability.

In Scenario 2 where the within-site variance is greater than the between-site variance, BIPmixed delivers the best predictive accuracy with the lowest MSE (2.320) and moderate variance (3.040). PCA2Step (2.822) provides the second-best MSE but exhibits the highest variance (5.265), indicating overfitting to specific views. BIP, while showing higher MSE (3.242), maintains a low FPR (0.138) and a perfect AUC (1.000). Cooperative Learning continues to excel in variance control (1.338) but suffers from an elevated FNR (0.924). 

In Scenario 3 where within-site variance is less than between-site, BIPmixed maintains its superiority with the lowest MSE (2.830) and moderate variance (2.596). BIP follows with an MSE of 3.213, outperforming PCA2Step (3.628) and Cooperative Learning (3.363). Variance patterns align with earlier scenarios, where Cooperative Learning retains the lowest variance (1.321). 

Overall, BIPmixed demonstrates superior predictive accuracy in more complex scenarios (2 and 3), while BIP shows competitive performance in a simpler setting (Scenario 1). PCA2Step serves as a non-integrative alternative but suffers from high variance, indicating less stability in its predictions. Cooperative Learning excels in variance control but struggles with feature selection in Scenarios 2 and 3, as reflected by its high FNR. These findings emphasize BIPmixed’s ability to capture hierarchically nested variability, achieving better prediction in such contexts. Consistently high AUC confirms strong feature selection, making it the most reliable method for multiview data similar to that studied.

\begin{figure}
    \centering
    \caption{Scatter and density plots of true \( y \) in red vs. predicted \( \hat{y} \) for three scenarios, colored by method: BIP, BIPmixed Cooperative Learning (Cooperative), and PCA 2-step (PCA2Step) from 1 replicate of each scenario. Red dashed line represents perfect calibration (slope = 1, intercept = 0) in scatter plots. In Scenarios 2 and 3, BIP and Cooperative depart from the 45-degree line, which is attributable to not accounting for random effect induced variance. BIPmixed, in contrast, is better calibrated in Scenarios 2 and 3, though PCA2Step reflects $y$'s variance.}
    \includegraphics[width=1.0\linewidth]{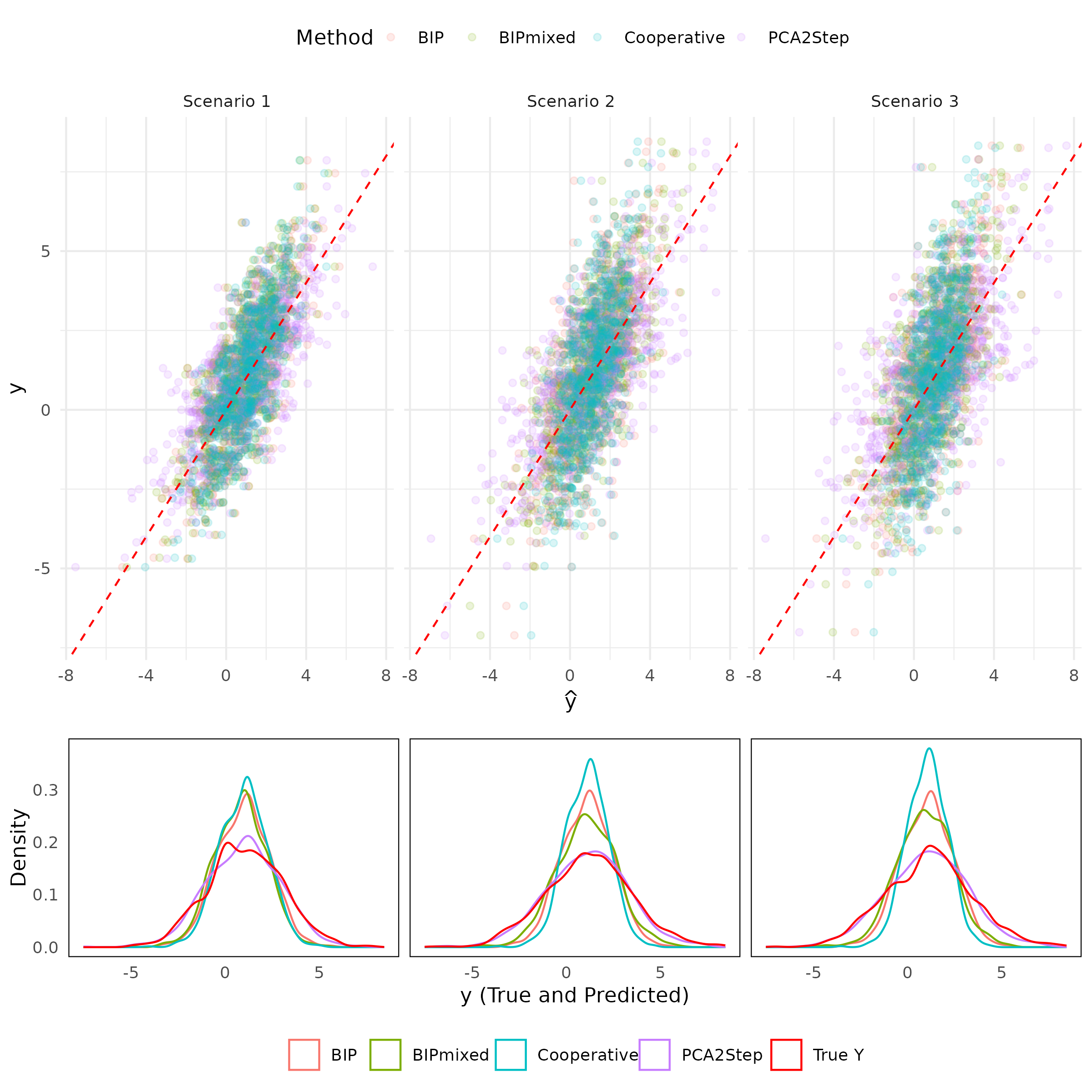}
    \label{fig:simulations-plot}
\end{figure}

\begin{table}[ht]
\centering
\caption{Performance metrics by Simulation Scenario (S.1, 2, 3) and Method (BIP, BIPmixed, Cooperative): Prediction metrics Mean Square Prediction Error (MSE), variance of predictions $\textrm{Var}(\hat{y})$, and feature selection metrics by Marginal Posterior Probability (MPP) averaged across views False Positive Rate (FPR), False Negative Rate (FNR; threshold = 0.5), and Area Under the Curve (AUC). MSE indicates prediction accuracy, $\text{Var}(\hat{y})$ reflects variability in predictions, and AUC measures discriminative ability by feature selection probability. For BIP and BIPmixed, AUC had a mean (SD) of 1.000 (0.000) across all scenarios. Feature selection metrics FPR, FNR, and AUC are not applicable (NA) to PCA2Step, and AUC to Cooperative. BIPmixed shows a lower MSE compared to all other methods in Scenario 2 (2.320) and Scenario 3 (2.830), indicating more accurate predictions in these cases. Second best MSE in Scenario 2: PCA2Step (2.822), and Scenario 3: BIP (3.213). \textbf{Bold} text indicates better performance for a given metric}

\setlength{\tabcolsep}{4pt}  
\makebox[\textwidth][c]{
\centering
\begin{tabular}{lll|lll}
  \hline
\textbf{Scenario - Method} & \textbf{MSE} & \textbf{Variance} & \textbf{FPR} & \textbf{FNR} \\ 
  \hline
1 - PCA2Step & 2.054 (0.084) & 4.091 (0.281) & NA & NA \\ 
  1 - BIP & \textbf{1.760} (0.075) & 2.133 (0.225) & 0.125 (0.176) & \textbf{0.000} (0.000) \\ 
  1 - BIPmixed & 1.816 (0.087) & 2.163 (0.234) & 0.131 (0.173) & \textbf{0.000} (0.000) \\ 
  1 - Cooperative & 1.826 (0.095) & \textbf{1.511} (0.201) & \textbf{0.000} (0.000) & 0.898 (0.021) \\ 
  \hline
  2 - PCA2Step & 2.822 (0.196) & 5.265 (0.539) & NA & NA \\ 
  2 - BIP & 3.242 (0.236) & 2.161 (0.271) & 0.138 (0.179) & \textbf{0.000} (0.000) \\ 
  2 - BIPmixed & \textbf{2.320} (0.131) & 3.040 (0.402) & 0.125 (0.157) & \textbf{0.000} (0.000) \\ 
  2 - Cooperative & 3.369 (0.289) & \textbf{1.338} (0.235) & \textbf{0.000} (0.000) & 0.924 (0.012) \\ 
  \hline
  3 - PCA2Step & 3.628 (0.204) & 5.210 (0.498) & NA & NA \\ 
  3 - BIP & 3.213 (0.198) & 2.202 (0.286) & 0.122 (0.165) & \textbf{0.000} (0.000) \\ 
  3 - BIPmixed & \textbf{2.830} (0.141) & 2.596 (0.319) & 0.144 (0.193) & \textbf{0.000} (0.000) \\ 
  3 - Cooperative & 3.363 (0.221) & \textbf{1.321} (0.193) & \textbf{0.000} (0.001) & 0.922 (0.024) \\ 
   \hline
\end{tabular}
}
\label{simulations-table}
\end{table}

\hypertarget{sec-discussion}{%
\section{Discussion}\label{sec-discussion}}

Simulations reveal performance differences in 4 methods—BIP, BIPmixed, Cooperative Learning, and PCA2Step—across scenarios that vary in random effects. When random effects are negligible (Scenario 1), BIP excels. BIPmixed and Cooperative Learning offer a comparable prediction, demonstrating flexibility even without a strong hierarchical signal. PCA2Step performs well but exhibits high $Var(\hat{y})$, indicating potential overfitting when not accounting for multiview structure. In nested scenarios (2 and 3) BIPmixed is the top performer in prediction and feature selection. BIP delivers reliable feature selection, with low false positive rates and perfect discrimination. Cooperative Learning has low prediction variance but sacrifices feature selection accuracy, as indicated by high FNR. 

From the ABCD Study data analysis, we found a combination of imaging and ELA features associated with externalizing problems. The selected ELA factors are consistent with findings from studies on borderline personality disorder and post-traumatic stress disorder, two trauma-based mental health conditions that may exhibit externalization of behaviors \citep{jowett_multiple_2020}, demonstrating the merits of BIPmixed. Furthermore, the superior prediction performance of BIPmixed when incorporating covariates directly into the outcome model highlights two plausible advantages. First, covariates may carry predictive information that is masked or diluted when treated as a separate view in the latent factor estimation process, aligning with principles seen in wide \& deep architectures \citep{cheng_wide_2016}. Analogous to the “wide” portion of such models, fixed and random effects in the outcome model leverage highly predictive covariates to enhance performance, complementing the ``deep" component of latent factor estimation. Second, the inclusion of covariates as fixed effects may yield efficiency gains in parameter estimation, reducing prediction variance. This aligns with the concept of efficiency augmentation described in recent literature, which suggests that explicitly modeling covariates improves the precision of latent factor estimates \citep{huling_subgroup_2018}. These findings underscore the benefits of integrating covariates into the outcome model to mitigate added variance from parameter estimation, emphasizing their critical role in enhancing prediction accuracy and stability.

These findings suggest that the accommodation of random effects can support prediction performance in large cohort study settings (and others involving hierarchical observations), though exploration to the extent that these random effects are not accounted for or muddled by already observed data is worth pursuing. Regardless, feature selected by BIPmixed can be interpreted in alignment with a priori known relationships between and within nested observations. Also, the inclusion of covariates in BIPmixed's outcome model can be helpful in situations where there are strongly predictive covariates available, though this has not been thoroughly investigated in this study and is likely of less utility in behavioral outcome prediction where signal-to-noise tends to be less than, for example, propensity to download mobile apps in the app store \citep{cheng_wide_2016}. Another benefit of BIPmixed in results not shown here is that accounting for random effects can result in better prediction at the tails of the random effect distribution relative to BIP, allowing especially for improvement of outcome predictions for sites or families that are far from the grand mean. In other words, families that are substantially dissimilar from the average outcome, i.e. are in the tails of the random effect distribution, experience less bias in prediction by BIPmixed, our method that accounts for family-wise heterogeneity. This helps combat a general limitation of machine learning methods, which is that they are usually optimized for the central tendency, resulting in predictions that perform less well for certain groups \citep{malik_hierarchy_2020}. 

Limitations include that in simpler situations or ones in which the random effects are not explicitly connected to the outcome $y$, BIPmixed has added variance in its predictions, which can reduce prediction performance. Furthermore, a general limitation of the BIP or BIPmixed framework is that the outcome is treated like just another view, which challenges its representation in less signal to noise rich settings (e.g. behavioral outcomes) or in more high-dimensional ones. Moreover, although BIPmixed can incorporate longitudinal data as an additional level in the hierarchical structure (i.e. multiple observations over time nested in an individual), temporal correlation may be distinct from the implied exchangeable correlation. BIPmixed has not been evaluated in the analysis and prediction of simulated or real longitudinal data, which is worth exploring. Given the importance of large-scale cohort data like that of the ABCD Study, addressing the aforementioned limitations will improve the benefits of multi-view analyses methods like BIPmixed. 

\hypertarget{sec-conc}{%
\section{Conclusion}\label{sec-conc}}

In conclusion, our study demonstrates the strengths and limitations of 4 multiview learning methods—BIP, BIPmixed, Cooperative Learning, and PCA2Step—across varied scenarios. BIP excels in settings with negligible random effects, while BIPmixed stands out in nested hierarchical contexts for both prediction accuracy and feature selection, leveraging covariates effectively in its outcome model. This highlights the importance of integrating covariates directly into the outcome model to reduce variance and enhance predictive performance, especially for data with hierarchical structures, as seen in the ABCD Study. However, BIPmixed's benefits diminish in simpler or low signal-to-noise settings, and its reliance on exchangeable correlation structures limits its utility for longitudinal data. Additionally, its treatment of the outcome as another view can challenge its effectiveness in high-dimensional or less informative settings. These findings underscore the need to refine methods like BIPmixed to address limitations, particularly for complex, large-scale cohort data, to further enhance prediction accuracy, feature interpretability, and adaptability across diverse data structures.

\hypertarget{disclosure}{%
\section{Disclosure}\label{disclosure}}

Authors declare they have no conflicts of interest to disclose. ChatGPT 4o was used for language improvement and coding assistance with rigorous revision and rearrangement.



\hypertarget{data-availability}{%
\section{Data Availability}\label{data-availability}}

Data used in the preparation of this article were obtained from the Adolescent Brain Cognitive Development (ABCD) Study (\urlstyle{https://abcdstudy.org}), held in the NIMH Data Archive (NDA). This is a multisite, longitudinal study designed to recruit more than 10,000 children age 9–10 and follow them over 10 years into early adulthood. The ABCD Study® is supported by the National Institutes of Health and additional federal partners under award numbers U01DA041048, U01DA050989, U01DA051016, U01DA041022, U01DA051018, U01DA051037, U01DA050987, U01DA041174, U01DA041106, U01DA041117, U01DA041028, U01DA041134, U01DA050988, U01DA051039, U01DA041156, U01DA041025, U01DA041120, U01DA051038, U01DA041148, U01DA041093, U01DA041089, U24DA041123, U24DA041147. A full list of supporters is available at https://abcdstudy.org/federal-partners.html. A listing of participating sites and a complete listing of the study investigators can be found at \urlstyle{https://abcdstudy.org/consortium_members/}. ABCD consortium investigators designed and implemented the study and/or provided data but did not necessarily participate in the analysis or writing of this report. This manuscript reflects the views of the authors and may not reflect the opinions or views of the NIH or ABCD consortium investigators. The ABCD data repository grows and changes over time. The ABCD data used in this report came from Data Release Version 5.1.

\hypertarget{code-availability}{%
\section{Code Availability}\label{code-availability}}

Scripts for processing ABCD Study data and performing the data analysis and simulation studies in this manuscript are available at: github.com/XXX 










  \bibliography{bibliography.bib}

\end{document}